\documentclass[twocolumn,amsmath,amssymb,floatfix,prl]{revtex4-1}
\usepackage{amsmath}
\usepackage{amssymb}
\usepackage{graphicx}% Include figure files

\def\Rb87{^{87}\rm{Rb}}					% Rb 87

\newcommand{\ket}[1]{\left|#1\right>}

\begin{document}
\title{Tunable Spin-Orbit Coupling via Strong Driving in Ultracold Atom Systems}

\author{K.~Jim\'{e}nez-Garc\'{\i}a$^{1,2}$}
\author{L.~J.~LeBlanc$^1$}
\author{R.~A.~Williams$^1$}
\author{M.~C.~Beeler$^1$}
\author{C.~Qu$^3$}
\author{M.~Gong$^3$}
\author{C.~Zhang$^3$}
\author{I.~B.~Spielman$^1$}
\email{ian.spielman@nist.gov}
\affiliation{$^1$Joint Quantum Institute, National Institute of Standards and Technology, and University of Maryland, Gaithersburg, Maryland, 20899, USA}
\affiliation{$^2$Departamento de F\'{\i}sica, Centro de Investigaci\'{o}n y Estudios Avanzados del Instituto Polit\'{e}cnico Nacional, M\'{e}xico D.F., 07360, M\'{e}xico}
\affiliation{$^3$Department of Physics, the University of Texas at Dallas, Richardson, TX,
75080 USA}

\date{\today}

\begin{abstract}
Spin-orbit coupling (SOC) is an essential ingredient in topological materials, conventional and quantum-gas based alike.~Engineered spin-orbit coupling in ultracold atom systems --unique in their experimental control and measurement opportunities-- provides a major opportunity to investigate and understand topological phenomena.~Here we experimentally demonstrate and theoretically analyze a technique for controlling SOC in a two component Bose-Einstein condensate using amplitude-modulated Raman coupling.
\end{abstract}
\maketitle

The properties of electronic materials are deeply entwined with their bandstructure --or more generally, their single-particle spectrum-- which gives rise to: conductors, semiconductors, conventional insulators and now topological insulators~\cite{Hasan2010}.~Understanding and controlling bandstructure in new ways therefore allows access to new phenomena.~Spin-orbit coupling (SOC) plays a fundamental role in most topological materials, linking the spin and the momentum of quantum particles.~The introduction of time-periodic perturbations to topologically trivial systems (quantum wells, solid-state materials, and ultracold atoms) can drive phase transitions to new ``Floquet topological phases"~\cite{Linder2011,Jotzu2014}.~For example, Floquet topological insulators arise from topologically trivial materials with spin-orbit coupling through time-periodic perturbations~\cite{Linder2011}.

In such materials, topological properties are induced and controlled by periodically modulating various terms in the single particle Hamiltonian.~In ultracold atom systems we precisely design, introduce and manipulate SOC by coupling the internal atomic degrees of freedom with laser fields~\cite{Dalibard2010}.~Here, we illuminated an ultracold atom system with a pair of ``Raman" lasers, inducing SOC in an effective two-level system~\cite{LinSOC_2011, Spielman2009,Cheuk2012,Wang2012,Zhang2012a,Hamner2014} with SOC strength defined by the laser geometry alone.~In this letter, we experimentally show that strongly modulating the Raman coupling tunes the SOC strength, independently of geometry and in agreement with theory.

\begin{figure}
 \begin{center}
 \includegraphics[width=3.5in]{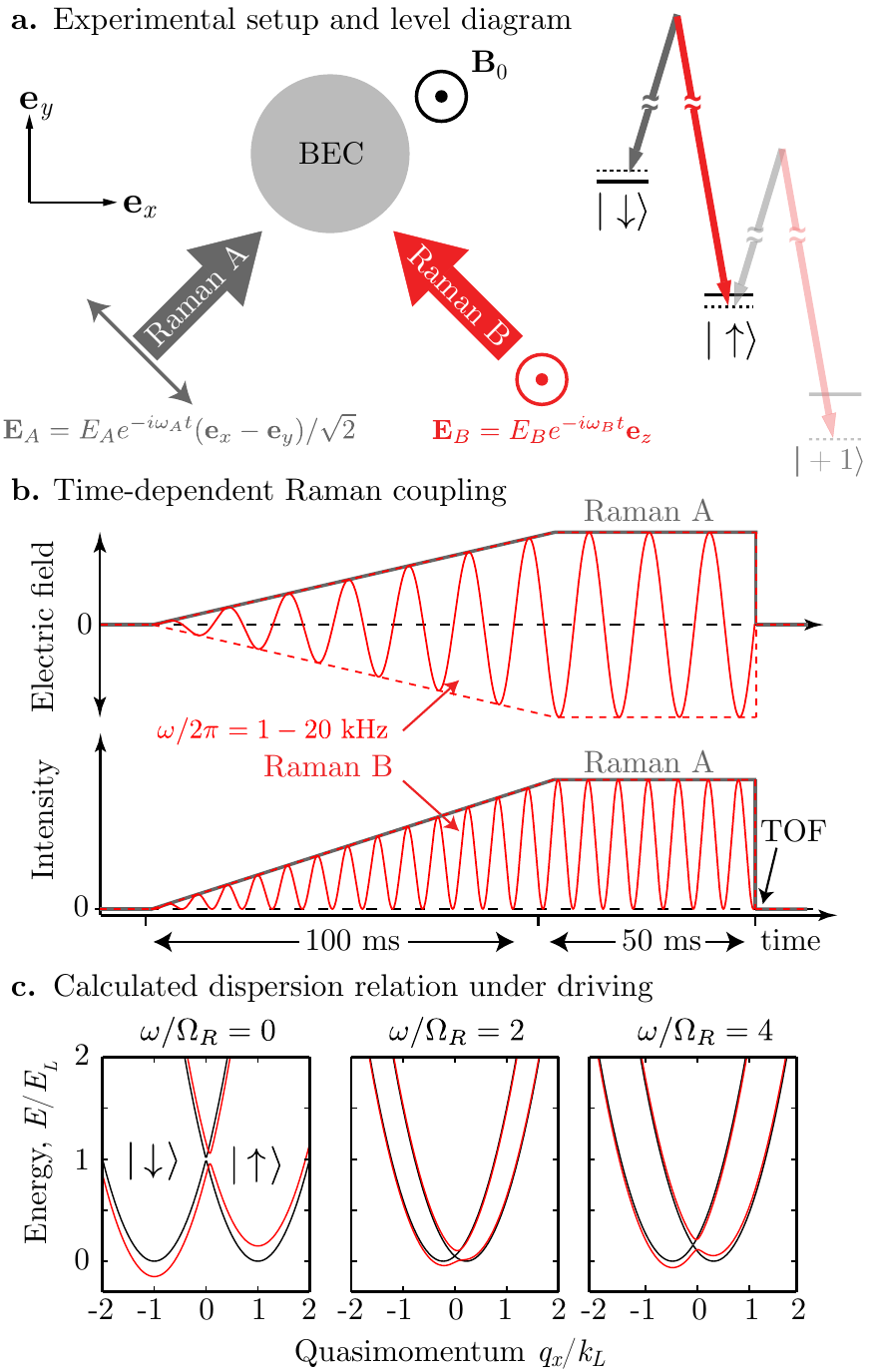}\\
 \end{center}
 \vspace{-0pt}
 \caption[Setup and level diagram.]{Setup and level diagram.~{\bf a.}~A uniform bias field $B {\bf e}_z$ Zeeman splits the hyperfine sublevels of an $f=1$~$\Rb87$~BEC, and a pair of Raman beams illuminate the atoms.~The field $B$ generates a large quadratic Zeeman shift $\hbar \epsilon/E_L \gg 1$ which effectively decouples the third spin state.~By adjusting the detuning we select the states $\ket{-1}=\ket{\downarrow}$ and $\ket{0}=\ket{\uparrow}$ to form an effective two-level system.~{\bf b.}\;Schematic of electric field and associated intensity ramps used in experiment~\cite{Juzeliunas2012} to modulate the Raman coupling strength $\Omega(t)$.~{\bf c}~Calculated dispersion relations from the time-periodic single particle Hamiltonian.~Black curves indicate: $\delta = 0$, and $\Omega_0 = 0$; while red curves indicate: ${\delta = -0.3 E_L}$, and $\Omega_0 = 0.1 E_L$.}
 \label{fig:SOC_ExpSetup}
\end{figure}

We engineered SOC in an effective two-level atom in a uniform magnetic field $B\hat{\bf e}_z$ that Zeeman-split the energy levels by $\hbar \omega_Z=g_F \mu_B B$, where $\mu_B$ is Bohr's
magneton and $g_F$ is the Land{\'e} $g$-factor.~These levels were coupled by a pair of orthogonally polarized Raman laser beams with angular frequencies $\omega_L$ and $\omega_L + \Delta \omega$ and a relative phase, as shown in Fig.~\ref{fig:SOC_ExpSetup}.~The lasers' frequency difference $\Delta \omega$ was set near $\omega_{Z}$, naturally defining an experimentally tunable detuning $\delta_0 = \Delta \omega - \omega_{Z}$.

In our experiment, we selected as our two-level system the $\ket{m_F = 0, -1} \equiv \ket{ \uparrow, \downarrow}$ hyperfine states of the $5S_{1/2}$, $f=1$ manifold of $^{87}\rm Rb$~\cite{LinSOC_2011}.~The Raman laser field coupled spin states $ {\ket{\downarrow, q_x = k_x - k_L}}$ to ${\ket{\uparrow, q_x = k_x + k_L}} $, differing in momentum by $2k_L$, where $q_x = k_x \pm k_L$ denotes the quasimomentum.~The recoil momentum ${k_L=2\pi \sin(\theta/2) / \lambda}$ and energy ${E_L = \hbar^2 k_L^2 / 2 m}$ set the relevant momentum and energy scales for Raman lasers intersecting at an angle $\theta$; here $\lambda$ is the laser wavelength and $m$ is the atomic mass.~In this experiment ${\theta = \pi /2}$, as shown in Fig.~\ref{fig:SOC_ExpSetup}~\footnote{For this effective two-level system we redefine the detuning as $\delta = \delta' + \epsilon$, to account for the quadratic contribution to the Zeeman shift, $\epsilon$~\cite{LinSOC_2011}.}.

In the frame rotating at angular frequency $\Delta\omega$ and after making the rotating wave approximation, the Hamiltonian combining both the kinetic and Raman coupling contributions is~\cite{LinSOC_2011} \begin{equation}\label{eqn:SOCHam_2x2}
\hat{H} = \left(\frac{\hbar^2 q_x^2 }{2m} + E_L\right) \hat{1} + \frac{\hbar \Omega}{2}\hat{\sigma}_x + \frac{\hbar \delta_0}{2}\hat{\sigma}_z + \alpha_0 q_x\hat{\sigma}_z,
\end{equation}
where~${\Omega \propto E^*_A E_B}$ is the Raman coupling~strength, $\hat{\sigma}_{x,y,z}$ are the Pauli matrices, and $E_{A,B}$ are the complex-valued optical electric field strengths (Fig.~\ref{fig:SOC_ExpSetup}a).~The last term describes SOC $-{\rm an~equal~sum~of~Rashba~and~Dresselhaus~SOC-}$~with strength $\alpha_0=2E_L / k_L\propto k_L$.~The resulting energy bands of the laser dressed atomic system $E_{\pm}(q_x)$ are obtained by diagonalizing ${\hat H}$ as a function of the quasimomentum $q_x$.~We focused atoms in the lowest energy band, where they experienced the energy-momentum dispersion relation given by $E_{-}(q_x)$.

The SOC strength ${\alpha_0\propto \sin(\theta/2)}$ depends only on the momentum difference between the Raman laser beams, reaches its maximum for counter-propagating beams, $\theta = \pi$.~Here we demonstrate a method for tuning the SOC strength $\alpha_0$ in real time: modulating the coupling strength $\hbar\Omega$ by controlling the intensity and phase of the Raman lasers~\cite{Juzeliunas2012,Zhang2011,Balaz2012, Struck2012}.~For rapid drive of the form $\Omega(t) = \Omega_0 + \Omega_R \cos(\omega t)$, and $\hbar \omega \gg 4 E_L$ the effective Floquet Hamiltonian retains the form of Eq.~\ref{eqn:SOCHam_2x2} but with renormalized coefficients $\Omega = \Omega_0$, $\delta = J_0 (\Omega_R / \omega) \delta_0$ and $\alpha = J_0 (\Omega_R / \omega) \alpha_0$.~$J_0$ is the zeroth order Bessel function of the first kind; i.e., $\alpha$ is an oscillatory function of $\Omega_0 / \omega$, generally decreasing in amplitude as $\Omega_0 / \omega$ increases~\cite{Zhang2011}.~

Our experiments started with nearly pure $\Rb87$ BECs in a crossed optical dipole trap, with frequencies $(f_x, f_y, f_z)=(32, 37, 100)~$Hz.~Prior to dressing the atoms with the Raman lasers, we prepared these BECs either in the spin state $\ket{\downarrow}$, $\ket{\uparrow}$, or an equal superposition thereof.~The $B=2.142~{\rm mT}$ bias field Zeeman split the $\ket{\uparrow}$ and $\ket{ \downarrow }$ states by $ \omega_Z /2\pi \approx 15~$MHz, detuning the unwanted $\ket{m_F = +1}$ state by $36 E_L$ from resonance.

\begin{figure}
 \begin{center}
 \includegraphics[width=3.5in]{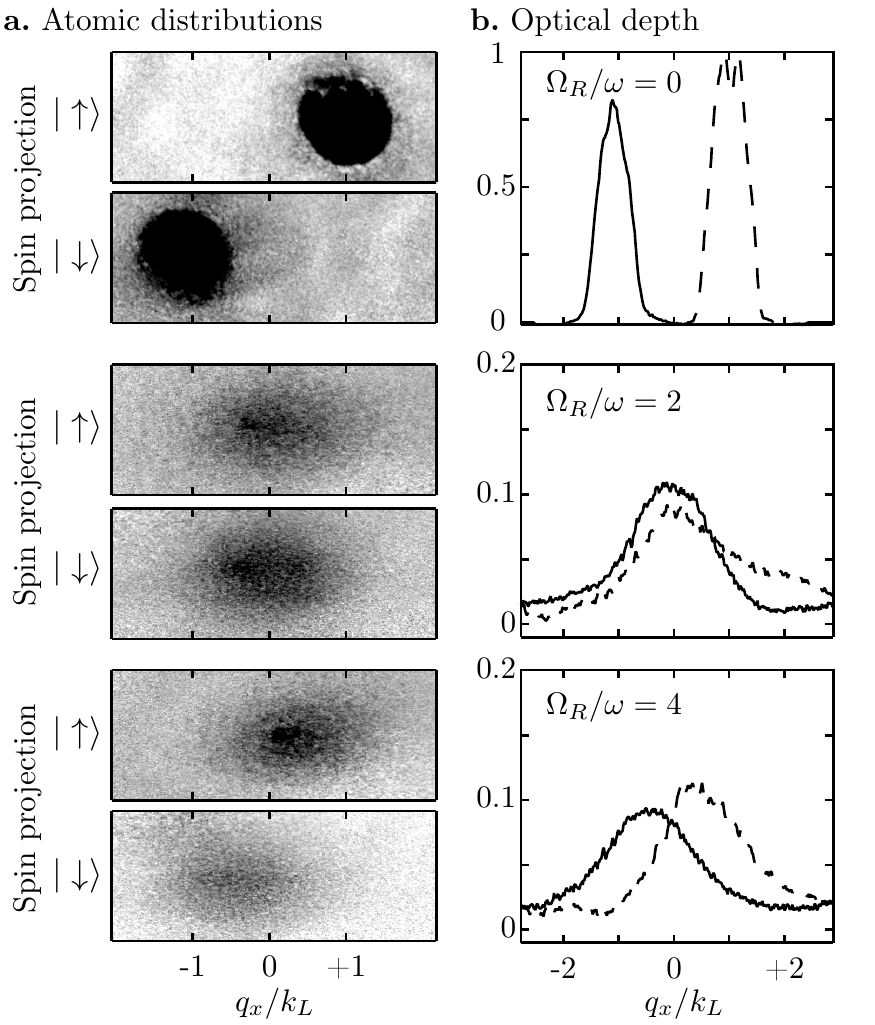}\\
 \end{center}
 \vspace{-0pt}
 \caption[OD Images.]{Absorption imaged TOF density distributions for $\omega/2\pi = 10~{\rm kHz}$.~The top, middle and bottom panels correspond to $\Omega_R/\omega=0, 2$ and 4, respectively.~{\bf a.}~(Top) When $\Omega_R/\omega=0$ the spin states are maximally separated by $\Delta k_x = 2k_L$.~(Middle)~When $\Omega_R/\omega=2$ the quasimomentum separation is practically zero.~(Bottom) When $\Omega_R/\omega=4$ the spin states again separate in quasimomentum.~{\bf b.}~The continuous (dashed) lines correspond to the optical depth, integrated along ${\bf e}_y$, for $\ket{\downarrow}$ ($\ket{\uparrow}$).}
 \label{fig:ODimages}
\end{figure}

\begin{figure}
 \begin{center}
 \includegraphics[width=3.5in]{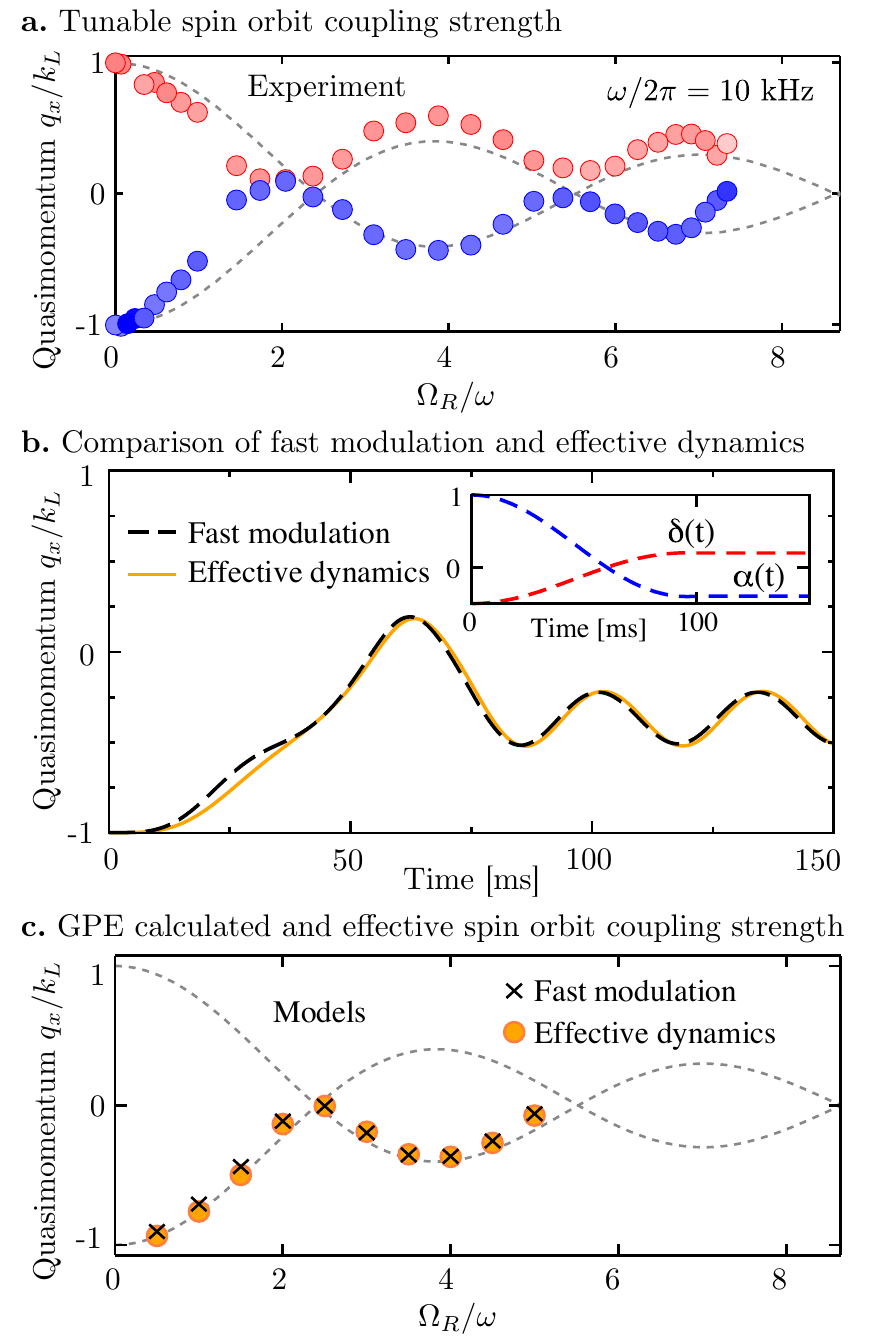}\\
 \end{center}
 \vspace{-0pt}
 \caption[Tunable SOC strength.]{Tunable SOC strength.~{\bf a.}~Measured (top) and computed (bottom) quasimomentum of systems driven at $\omega /2\pi = 10~$kHz as a function of $\Omega_R/\omega$.~The initial state was an equal superposition of spin states.~The blue (red) symbols represent the quasimomentum of the $\ket{\downarrow}$ ($\ket{\uparrow}$) atoms.~{\bf b.}~Dynamics of the BEC's quasimomentum for $\Omega_R/\omega = 4$.~The dashed and continuous curves correspond to: fast modulation of $\Omega(t)$ following the exact experiment; and to the effective dynamics predicted from the resulting effective  $\alpha(t)$ and $\delta(t)$ (shown in the inset), respectively.~{\bf c.}~Calculated mean value of the quasimomentum after the laser ramp-up period.~This simulation shows a ``reflection" of the quasimomentum at $q_x = 0$ for constant $\hbar \Omega_0 = 0.3 E_L$, $\hbar \delta_0 = -0.5E_L$ and $\omega/2\pi = 20\;{\rm KHz}$.~The dashed curves in panels {\bf a} and {\bf c} correspond to the Bessel function $J_0(\Omega_R/\omega)$.}
 \label{fig:TunableSOC}
\end{figure}

We optically dressed the atoms with a pair of $\lambda = 790.1$~nm Raman lasers propagating along ${\bf e}_y \pm {\bf e}_x$ (Fig.~\ref{fig:SOC_ExpSetup}), and controlled $\delta_0$ by making small changes to $B$.~The Raman coupling strength $\hbar \Omega$ was experimentally controlled by the intensity of the lasers, and we inverted the sign of $\Omega$ by shifting the beams' relative phase by $\pi$~\cite{Juzeliunas2012}.~Each Raman beam (labeled A and B, respectively) was ramped from zero to its final intensity in 100~ms following a linear envelope; however, the intensity of Raman laser B was additionally modulated sinusoidally~(Fig.~\ref{fig:SOC_ExpSetup}b).~The atoms were then held for 50~ms, after which all potentials were turned-off.~The atomic ensemble expanded for a 34.45~ms time-of-flight (TOF) before absorption imaging.~Using a magnetic field gradient  during part of TOF we separated the spin components along~${\bf e}_y$.

We determined the SOC strength from direct measurements of atomic momentum distributions as shown in Fig.~\ref{fig:ODimages}.~We first studied systems driven at $\omega/2\pi = 10~$kHz and 20~kHz and found momentum distributions in excellent agreement with the expected behavior, i.e.~the atoms adiabatically followed degenerate ground states of the driven Raman Hamiltonian located at $q_x = \pm(\alpha/\alpha_0) k_L$ as we tuned $\Omega_R/\omega$.~Figure~\ref{fig:TunableSOC}a constitutes the main result of our work and demonstrates experimental control on the SOC strength $\alpha(\Omega_R/\omega)$, for systems driven at $\omega/2\pi = 10~$kHz.~As $\Omega_R/\omega$ increased, we observed the Bessel-function behavior of $\alpha$.

We compared our data to the simulated dynamics of the BEC, governed by the time-dependent Gross-Pitaevskii equation (TDGPE)
\begin{equation}
i\hbar \frac{\partial \Psi}{ \partial t} = \left [ H(t)+V({\bf r}) + H_{I}\right ] \Psi , \label{TGP}
\end{equation}%
where ${\Psi =(\Psi _{\downarrow },\Psi _{\uparrow })^{T}}$ is a two-component wave function.~We numerically simulated 3D~BECs with ${N=10^5}$ atoms in a harmonic confining potential~${V\left( \mathbf{r}\right) = m(\omega _{x}^{2}x^{2}+\omega _{y}^{2}y^{2}+\omega _{z}^{2}z^{2})/2}$, and with atomic density-density interactions described by $H_{I}= {\rm diag}\left( g_{\uparrow \downarrow }|\Psi _{\uparrow}|^{2}+g_{\downarrow \downarrow }|\Psi _{\downarrow }|^{2},g_{\uparrow \uparrow }|\Psi _{\uparrow }|^{2}+g_{\uparrow \downarrow }|\Psi _{\downarrow}|^{2}\right)$.~The interaction constants $g_{\downarrow \downarrow }=g_{\uparrow \downarrow }=4\pi \hbar^{2}N(c_{0}+c_{2})/m$ and $g_{\uparrow \uparrow }=4\pi \hbar^{2}Nc_{0}/m$ are derived from $^{87}{\rm Rb}$'s \textit{s}-wave scattering lengths ${c_{0}=100.86a_{\rm B}}$, ${c_{2}=-0.46a_{\rm B}}$ ($a_{\rm B}$ is the Bohr radius).~We obtained the ${t=0}$ initial state (before modulation) using imaginary time-evolution the BEC initially polarized in one spin component, and then explicitly time-evolved with the TDGPE~\cite{Bao2003} including the full experimental time-dependent Raman coupling $\Omega\left( t\right)$.

\begin{figure}
 \begin{center}
 \includegraphics[width=3.5in]{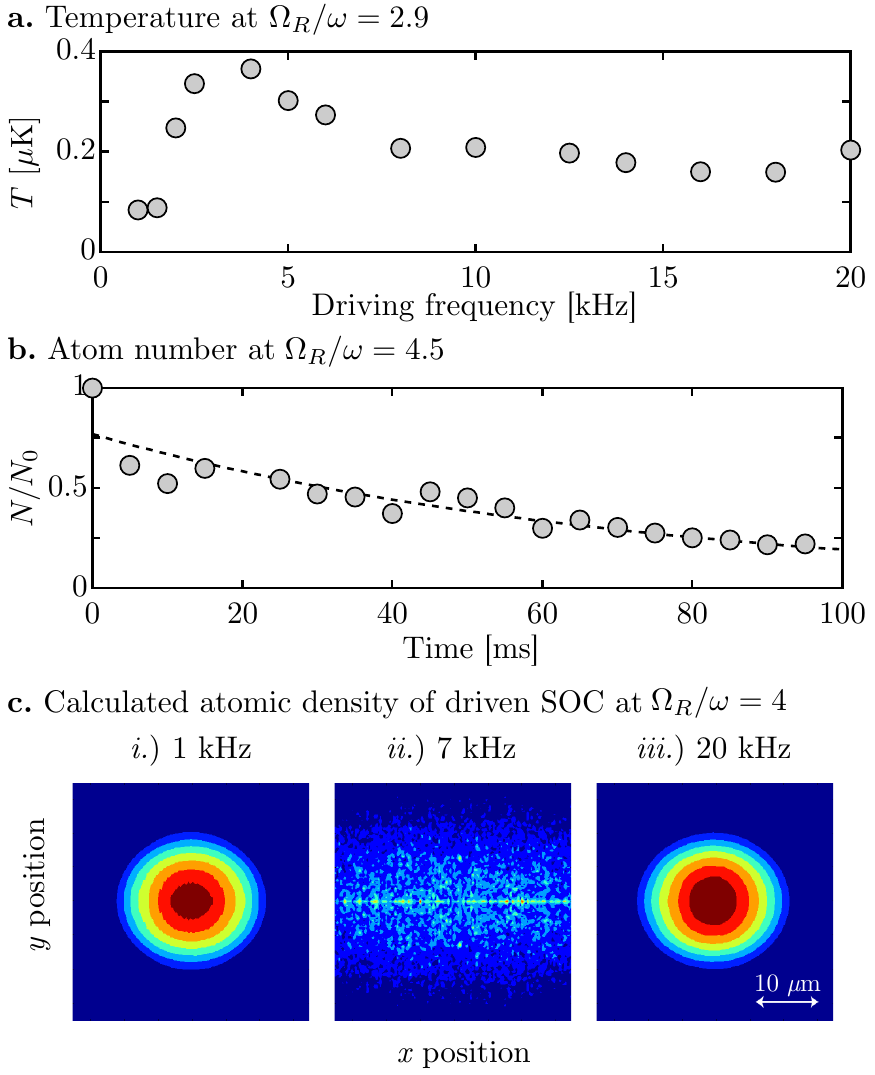}\\
 \end{center}
 \vspace{-0pt}
 \caption[Heating.]{Heating and loss in the optically driven SOC system.~{\bf a.} Temperature of the atomic sample measured from the thermal fraction of the momentum distribution as a function of $\omega$ at ${\Omega_R/\omega=2.9}$.~{\bf b.} Number of atoms remaining in the driven system for ${\Omega_R/\omega=4.5}$, an exponential fit gave a 1/e lifetime $\tau = 72.3$~{ms}.~{\bf c.}~Density distribution $\left\vert \Psi _{\downarrow }\right\vert^{2} $ of the BEC for different modulation frequencies $\omega/2\pi$ at ${\Omega_R/\omega=4}$.~Panels {\it i.}) and {\it iii.}) correspond to the stable region, while panel {\it ii.}) shows the unstable region in the GP simulation arising from the coupling to higher energy bands and nonlinear interactions.}
 \label{fig:Heating}
\end{figure}

We modeled our experimental results under two relevant schemes.~{\it Fast modulation} corresponds to the time evolution of the GPE explicitly including the the full modulated Raman coupling $\Omega(t)$ (as in the experiment).~{\it Effective dynamics} describes simulations that instead used the Bessel function modified effective parameters $\alpha(t)$ and $\delta(t)$ in which $\alpha(t)$ and  $\delta(t)$ were slowly ramped as the Raman lasers turned on (illustrated in Fig.~\ref{fig:TunableSOC}b, inset).~Figure~\ref{fig:TunableSOC}b shows the time evolution of the quasimomentum and demonstrates that the effective description is in good agreement with the explicit fast modulation simulation.~Oscillations of the BEC's quasimomentum around the local band minimum (for ${t>100~{\rm ms}}$) result from imperfect adiabaticity during the ramp-up process of the Raman lasers.~Figure~\ref{fig:TunableSOC}c displays the final quasimomentum averaged over one $\approx 25\ {\rm ms}$ oscillation period, giving to the band minimum.~In experiment, we found that the system very rapidly relaxed to the local band minimum; therefore the measured quasimomentum shown in Fig.~\ref{fig:TunableSOC}a.

We experimentally observed that as $J_0(\Omega_R/\omega)$ became negative, the individual spins did not pass through $q_x=0$ as might be expected, but rather ``reflected'' and continued following the Bessel envelope without changing sign (Fig.~\ref{fig:TunableSOC}a); this was the case for both the spin superposition and the single spin data.~Our simulations show that this reflection is present when the Raman coupling offset $\hbar\Omega _{0}$ and detuning $\hbar \delta_0$ are small but non-zero.~For the simulation shown in Fig.~\ref{fig:TunableSOC}c, we used $\hbar\Omega_0 = 0.3E_L$ and $\hbar\delta_0 = -0.5EL$; furthermore, using $\hbar\delta_0 = -0.4E_L$ and  $-0.2E_L$, the quasimomentum also displayed a reflection at $q_x=0$.~However, for $\delta_0 = 0$ the ``reflection" depended on the actual value of the Raman coupling and driving frequency.~

Physically, when these terms are small the atoms undergo a Majorana spin-flip as $J_0(\Omega_R/\omega)$ changes sign but are able to adiabatically follow when they are non-zero.~Because $\alpha, \delta\propto J_0(\Omega_R / \omega)$ change sign simultaneously, the $q_x$ for the local minima in $E_{-}(q_x)$ associated with each spin state do not change sign, so the local minima reflect from $q_x=0$.~This argument can also be understood by considering the red curves in Fig.~\ref{fig:SOC_ExpSetup}c, showing a progression of effective SOC dispersion relations with nonzero $\Omega_0$ and $\delta$; it is evident that atoms which start in the lower (left) minimum will stay in that minimum even after the minima have merged and separated once more.

In the experiment, $\Omega(t)$ was determined by the intensity and relative phase of two Raman lasers as controlled by acousto-optic modulators (AOMs); it is likely that a small DC contribution to the AOM's drive gave $\Omega(t)$ a small non-zero average at the 5\% level.~In the strong driving region, $\hbar \Omega _{R}>\hbar\omega \approx10E_{L}$, this corresponds to a $\hbar \Omega _{0}\sim 0.5E_{L}$ offset.~Furthermore, in our experiment, small detunings $\hbar |\delta_0|\sim0.1E_{L}$ were generally present.

%up is m_F = 0
%down is m_F = -1

In addition, Fig.~\ref{fig:ODimages} shows that the system is heated in the presence of the drive.~Figure~\ref{fig:Heating}a parameterizes this effect in terms of the temperature of the driven system as a function of driving frequency at fixed $\Omega_R/\omega = 2.9$.~The heating was most pronounced in the range $2~{\rm kHz}<\omega/2\pi<7~{\rm kHz}$ and reached a plateau for $\omega/2\pi>10~{\rm kHz}$.~Because our atoms are continuously evaporating from the shallow optical dipole trap, this heating drives rapid atom loss, as plotted in Fig.~\ref{fig:Heating}b for $\Omega_R/\omega = 4.5$ and $\omega/2\pi=10~{\rm kHz}$.

This unwanted heating is present as dephasing in our zero-temperature GPE model and described in terms of the BEC's stability under driving (Fig.~\ref{fig:Heating}c).~The BECs stability depends on the modulation frequency.~For very large $\omega $, the time-dependent terms average out (rotating wave approximation), and the dynamics of the BEC follow the effective modulation without instability.~For very small $\omega $, the SOC strength barely changes and there is no instability.~In the intermediate regime, the strong instability of the modulated BEC is observed because the modulation effectively couples to the BEC's collective modes.

The unstable range of drive frequencies is larger for stronger interactions, and vanishes for vanishing interactions where the simple effective quasi-eigenstates become exact.~In the unstable region, the BEC is destroyed after a modulation time of just $10-50 $~ms, at which point the momentum space distribution is dominated by high momentum excitations, which would be interpreted as thermal excitations observed in experiment.~Figure~\ref{fig:Heating}c plots the GPE-computed density distributions in stable and unstable regions for $J_{0}(\Omega _{R}/\omega )=-0.4$.~For this simulation, we found frequencies in the range $2\ {\rm kHz}$ $\lesssim \omega/2\pi \lesssim 15\ {\rm kHz}$ gave unstable behavior.~For $\omega/2\pi \lesssim2\ {\rm kHz}$, the BEC is stable but the simple Bessel function description is not valid.~For $\omega/2\pi \gtrsim 15\ {\rm kHz}$ the system is stable and Bessel function description applies.~In general, the exact size of the unstable region depends on both $\Omega _{R}/\omega$ and interaction strength; thus while the experimental data falls into the unstable region, the exact location of the boundary may be influenced by the constant atom number ($N=10^5$) used in the numerical simulation.~For this experiment we avoided the larger $\omega$'s required to enter the stable regime because this also requires larger $\Omega _{R}$, leading to unwanted spontaneous-emission driven heating (not included in our GPE calculation).

Here, we demonstrated a technique to control the coupling strength in a light-induced SOC system.~Our technique relies on modulating the Raman laser field illuminating an ultracold atom system.~The measured SOC strength as a function of the dimensionless Raman coupling strength $\Omega_R / \omega$, is in good agreement with theory.~This work shows that Raman modulation is a powerful way to control SOC in quantum gases, in analogy to modulated lattice experiments~\cite{Lignier2007,Struck2012}.

We appreciate enlightening conversations with G.~Juzeliunas, N.~R.~Cooper, and W.~D.~Phillips; additionally we thank Dina Genkina for carefully reading our manuscript.~This work was partially supported by the ARO with funding from DARPA's OLE program and the Atomtronics-MURI; and the NSF through the JQI Physics Frontier Center.~K.J.-G.~thanks CONACYT; L.J.L.~thanks NSERC; M.C.B.~thanks the NIST-ARRA program and C.Q., M.G.~and C.Z.~are supported by ARO (W911NF-12-1-0334), AFOSR (FA9550-13-1-0045), and NSF-PHY (1104546).

\bibliography{tunableSOC}

\begin{thebibliography}{17}%
\makeatletter
\providecommand \@ifxundefined [1]{%
 \@ifx{#1\undefined}
}%
\providecommand \@ifnum [1]{%
 \ifnum #1\expandafter \@firstoftwo
 \else \expandafter \@secondoftwo
 \fi
}%
\providecommand \@ifx [1]{%
 \ifx #1\expandafter \@firstoftwo
 \else \expandafter \@secondoftwo
 \fi
}%
\providecommand \natexlab [1]{#1}%
\providecommand \enquote  [1]{``#1''}%
\providecommand \bibnamefont  [1]{#1}%
\providecommand \bibfnamefont [1]{#1}%
\providecommand \citenamefont [1]{#1}%
\providecommand \href@noop [0]{\@secondoftwo}%
\providecommand \href [0]{\begingroup \@sanitize@url \@href}%
\providecommand \@href[1]{\@@startlink{#1}\@@href}%
\providecommand \@@href[1]{\endgroup#1\@@endlink}%
\providecommand \@sanitize@url [0]{\catcode `\\12\catcode `\$12\catcode
  `\&12\catcode `\#12\catcode `\^12\catcode `\_12\catcode `\%12\relax}%
\providecommand \@@startlink[1]{}%
\providecommand \@@endlink[0]{}%
\providecommand \url  [0]{\begingroup\@sanitize@url \@url }%
\providecommand \@url [1]{\endgroup\@href {#1}{\urlprefix }}%
\providecommand \urlprefix  [0]{URL }%
\providecommand \Eprint [0]{\href }%
\providecommand \doibase [0]{http://dx.doi.org/}%
\providecommand \selectlanguage [0]{\@gobble}%
\providecommand \bibinfo  [0]{\@secondoftwo}%
\providecommand \bibfield  [0]{\@secondoftwo}%
\providecommand \translation [1]{[#1]}%
\providecommand \BibitemOpen [0]{}%
\providecommand \bibitemStop [0]{}%
\providecommand \bibitemNoStop [0]{.\EOS\space}%
\providecommand \EOS [0]{\spacefactor3000\relax}%
\providecommand \BibitemShut  [1]{\csname bibitem#1\endcsname}%
\let\auto@bib@innerbib\@empty
%</preamble>
\bibitem [{\citenamefont {Hasan}\ and\ \citenamefont {Kane}(2010)}]{Hasan2010}%
  \BibitemOpen
  \bibfield  {author} {\bibinfo {author} {\bibfnamefont {M.~Z.}\ \bibnamefont
  {Hasan}}\ and\ \bibinfo {author} {\bibfnamefont {C.~L.}\ \bibnamefont
  {Kane}},\ }\href {\doibase 10.1103/RevModPhys.82.3045} {\bibfield  {journal}
  {\bibinfo  {journal} {Rev. Mod. Phys.}\ }\textbf {\bibinfo {volume} {82}},\
  \bibinfo {pages} {3045} (\bibinfo {year} {2010})}\BibitemShut {NoStop}%
\bibitem [{\citenamefont {Linder}\ \emph {et~al.}(2011)\citenamefont {Linder},
  \citenamefont {Refael},\ and\ \citenamefont {Galitski}}]{Linder2011}%
  \BibitemOpen
  \bibfield  {author} {\bibinfo {author} {\bibfnamefont {N.~H.}\ \bibnamefont
  {Linder}}, \bibinfo {author} {\bibfnamefont {G.}~\bibnamefont {Refael}}, \
  and\ \bibinfo {author} {\bibfnamefont {V.}~\bibnamefont {Galitski}},\ }\href
  {\doibase 10.1038/nphys1926} {\bibfield  {journal} {\bibinfo  {journal}
  {Nature Phys.}\ }\textbf {\bibinfo {volume} {7}},\ \bibinfo {pages} {490}
  (\bibinfo {year} {2011})}\BibitemShut {NoStop}%
\bibitem [{\citenamefont {Jotzu}\ \emph {et~al.}(2014)\citenamefont {Jotzu},
  \citenamefont {Messer}, \citenamefont {Desbuquois}, \citenamefont {Lebrat},
  \citenamefont {Uehlinger}, \citenamefont {Greif},\ and\ \citenamefont
  {Esslinger}}]{Jotzu2014}%
  \BibitemOpen
  \bibfield  {author} {\bibinfo {author} {\bibfnamefont {G.}~\bibnamefont
  {Jotzu}}, \bibinfo {author} {\bibfnamefont {M.}~\bibnamefont {Messer}},
  \bibinfo {author} {\bibfnamefont {R.}~\bibnamefont {Desbuquois}}, \bibinfo
  {author} {\bibfnamefont {M.}~\bibnamefont {Lebrat}}, \bibinfo {author}
  {\bibfnamefont {T.}~\bibnamefont {Uehlinger}}, \bibinfo {author}
  {\bibfnamefont {D.}~\bibnamefont {Greif}}, \ and\ \bibinfo {author}
  {\bibfnamefont {T.}~\bibnamefont {Esslinger}},\ }\href {\doibase
  10.1038/nature13915} {\bibfield  {journal} {\bibinfo  {journal} {Nature}\
  }\textbf {\bibinfo {volume} {515}},\ \bibinfo {pages} {237} (\bibinfo {year}
  {2014})}\BibitemShut {NoStop}%
\bibitem [{\citenamefont {Dalibard}\ \emph {et~al.}(2011)\citenamefont
  {Dalibard}, \citenamefont {Gerbier}, \citenamefont
  {Juzeli\ifmmode~\bar{u}\else\=u\fi{}nas},\ and\ \citenamefont
  {\"Ohberg}}]{Dalibard2010}%
  \BibitemOpen
  \bibfield  {author} {\bibinfo {author} {\bibfnamefont {J.}~\bibnamefont
  {Dalibard}}, \bibinfo {author} {\bibfnamefont {F.}~\bibnamefont {Gerbier}},
  \bibinfo {author} {\bibfnamefont {G.}~\bibnamefont
  {Juzeli\ifmmode~\bar{u}\else\=u\fi{}nas}}, \ and\ \bibinfo {author}
  {\bibfnamefont {P.}~\bibnamefont {\"Ohberg}},\ }\href {\doibase
  10.1103/RevModPhys.83.1523} {\bibfield  {journal} {\bibinfo  {journal} {Rev.
  Mod. Phys.}\ }\textbf {\bibinfo {volume} {83}},\ \bibinfo {pages} {1523}
  (\bibinfo {year} {2011})}\BibitemShut {NoStop}%
\bibitem [{\citenamefont {Lin}\ \emph {et~al.}(2011)\citenamefont {Lin},
  \citenamefont {Jim{\'{e}}nez-Garc\'{i}a},\ and\ \citenamefont
  {Spielman}}]{LinSOC_2011}%
  \BibitemOpen
  \bibfield  {author} {\bibinfo {author} {\bibfnamefont {Y.-J.}\ \bibnamefont
  {Lin}}, \bibinfo {author} {\bibfnamefont {K.}~\bibnamefont
  {Jim{\'{e}}nez-Garc\'{i}a}}, \ and\ \bibinfo {author} {\bibfnamefont {I.~B.}\
  \bibnamefont {Spielman}},\ }\href {\doibase 10.1038/nature09887} {\bibfield
  {journal} {\bibinfo  {journal} {Nature}\ }\textbf {\bibinfo {volume} {471}},\
  \bibinfo {pages} {83} (\bibinfo {year} {2011})}\BibitemShut {NoStop}%
\bibitem [{\citenamefont {Spielman}(2009)}]{Spielman2009}%
  \BibitemOpen
  \bibfield  {author} {\bibinfo {author} {\bibfnamefont {I.~B.}\ \bibnamefont
  {Spielman}},\ }\href {\doibase 10.1103/PhysRevA.79.063613} {\bibfield
  {journal} {\bibinfo  {journal} {Phys. Rev. A}\ }\textbf {\bibinfo {volume}
  {79}},\ \bibinfo {pages} {063613} (\bibinfo {year} {2009})}\BibitemShut
  {NoStop}%
\bibitem [{\citenamefont {Cheuk}\ \emph {et~al.}(2012)\citenamefont {Cheuk},
  \citenamefont {Sommer}, \citenamefont {Hadzibabic}, \citenamefont {Yefsah},
  \citenamefont {Bakr},\ and\ \citenamefont {Zwierlein}}]{Cheuk2012}%
  \BibitemOpen
  \bibfield  {author} {\bibinfo {author} {\bibfnamefont {L.~W.}\ \bibnamefont
  {Cheuk}}, \bibinfo {author} {\bibfnamefont {A.~T.}\ \bibnamefont {Sommer}},
  \bibinfo {author} {\bibfnamefont {Z.}~\bibnamefont {Hadzibabic}}, \bibinfo
  {author} {\bibfnamefont {T.}~\bibnamefont {Yefsah}}, \bibinfo {author}
  {\bibfnamefont {W.~S.}\ \bibnamefont {Bakr}}, \ and\ \bibinfo {author}
  {\bibfnamefont {M.~W.}\ \bibnamefont {Zwierlein}},\ }\href {\doibase
  10.1103/PhysRevLett.109.095302} {\bibfield  {journal} {\bibinfo  {journal}
  {Phys. Rev. Lett.}\ }\textbf {\bibinfo {volume} {109}},\ \bibinfo {pages}
  {095302} (\bibinfo {year} {2012})}\BibitemShut {NoStop}%
\bibitem [{\citenamefont {Wang}\ \emph {et~al.}(2012)\citenamefont {Wang},
  \citenamefont {Yu}, \citenamefont {Fu}, \citenamefont {Miao}, \citenamefont
  {Huang}, \citenamefont {Chai}, \citenamefont {Zhai},\ and\ \citenamefont
  {Zhang}}]{Wang2012}%
  \BibitemOpen
  \bibfield  {author} {\bibinfo {author} {\bibfnamefont {P.}~\bibnamefont
  {Wang}}, \bibinfo {author} {\bibfnamefont {Z.-Q.}\ \bibnamefont {Yu}},
  \bibinfo {author} {\bibfnamefont {Z.}~\bibnamefont {Fu}}, \bibinfo {author}
  {\bibfnamefont {J.}~\bibnamefont {Miao}}, \bibinfo {author} {\bibfnamefont
  {L.}~\bibnamefont {Huang}}, \bibinfo {author} {\bibfnamefont
  {S.}~\bibnamefont {Chai}}, \bibinfo {author} {\bibfnamefont {H.}~\bibnamefont
  {Zhai}}, \ and\ \bibinfo {author} {\bibfnamefont {J.}~\bibnamefont {Zhang}},\
  }\href {\doibase 10.1103/PhysRevLett.109.095301} {\bibfield  {journal}
  {\bibinfo  {journal} {Phys. Rev. Lett.}\ }\textbf {\bibinfo {volume} {109}},\
  \bibinfo {pages} {095301} (\bibinfo {year} {2012})}\BibitemShut {NoStop}%
\bibitem [{\citenamefont {Zhang}\ \emph {et~al.}(2012)\citenamefont {Zhang},
  \citenamefont {Ji}, \citenamefont {Chen}, \citenamefont {Zhang},
  \citenamefont {Du}, \citenamefont {Yan}, \citenamefont {Pan}, \citenamefont
  {Zhao}, \citenamefont {Deng}, \citenamefont {Zhai}, \citenamefont {Chen},\
  and\ \citenamefont {Pan}}]{Zhang2012a}%
  \BibitemOpen
  \bibfield  {author} {\bibinfo {author} {\bibfnamefont {J.-Y.}\ \bibnamefont
  {Zhang}}, \bibinfo {author} {\bibfnamefont {S.-C.}\ \bibnamefont {Ji}},
  \bibinfo {author} {\bibfnamefont {Z.}~\bibnamefont {Chen}}, \bibinfo {author}
  {\bibfnamefont {L.}~\bibnamefont {Zhang}}, \bibinfo {author} {\bibfnamefont
  {Z.-D.}\ \bibnamefont {Du}}, \bibinfo {author} {\bibfnamefont
  {B.}~\bibnamefont {Yan}}, \bibinfo {author} {\bibfnamefont {G.-S.}\
  \bibnamefont {Pan}}, \bibinfo {author} {\bibfnamefont {B.}~\bibnamefont
  {Zhao}}, \bibinfo {author} {\bibfnamefont {Y.-J.}\ \bibnamefont {Deng}},
  \bibinfo {author} {\bibfnamefont {H.}~\bibnamefont {Zhai}}, \bibinfo {author}
  {\bibfnamefont {S.}~\bibnamefont {Chen}}, \ and\ \bibinfo {author}
  {\bibfnamefont {J.-W.}\ \bibnamefont {Pan}},\ }\href@noop {} {\bibfield
  {journal} {\bibinfo  {journal} {Phys. Rev. Lett.}\ }\textbf {\bibinfo
  {volume} {109}},\ \bibinfo {pages} {115301} (\bibinfo {year}
  {2012})}\BibitemShut {NoStop}%
\bibitem [{\citenamefont {Hamner}\ \emph {et~al.}(2014)\citenamefont {Hamner},
  \citenamefont {Qu}, \citenamefont {Zhang}, \citenamefont {Chang},
  \citenamefont {Gong}, \citenamefont {Zhang},\ and\ \citenamefont
  {Engels}}]{Hamner2014}%
  \BibitemOpen
  \bibfield  {author} {\bibinfo {author} {\bibfnamefont {C.}~\bibnamefont
  {Hamner}}, \bibinfo {author} {\bibfnamefont {C.}~\bibnamefont {Qu}}, \bibinfo
  {author} {\bibfnamefont {Y.}~\bibnamefont {Zhang}}, \bibinfo {author}
  {\bibfnamefont {J.}~\bibnamefont {Chang}}, \bibinfo {author} {\bibfnamefont
  {M.}~\bibnamefont {Gong}}, \bibinfo {author} {\bibfnamefont {C.}~\bibnamefont
  {Zhang}}, \ and\ \bibinfo {author} {\bibfnamefont {P.}~\bibnamefont
  {Engels}},\ }\href@noop {} {\bibfield  {journal} {\bibinfo  {journal} {Nature
  Communications}\ }\textbf {\bibinfo {volume} {5}} (\bibinfo {year}
  {2014})}\BibitemShut {NoStop}%
\bibitem [{Note1()}]{Note1}%
  \BibitemOpen
  \bibinfo {note} {For this effective two-level system we redefine the detuning
  as $\delta = \delta ' + \epsilon $, to account for the quadratic contribution
  to the Zeeman shift, $\epsilon $~\cite {LinSOC_2011}.}\BibitemShut {Stop}%
\bibitem [{\citenamefont {{Juzeli\ifmmode \bar{u}\else\=u\fi{}nas}}\ and\
  \citenamefont {Spielman}(2012)}]{Juzeliunas2012}%
  \BibitemOpen
  \bibfield  {author} {\bibinfo {author} {\bibfnamefont {G.}~\bibnamefont
  {{Juzeli\ifmmode \bar{u}\else\=u\fi{}nas}}}\ and\ \bibinfo {author}
  {\bibfnamefont {I.}~\bibnamefont {Spielman}},\ }\href {\doibase
  10.1088/1367-2630/14/12/123022} {\bibfield  {journal} {\bibinfo  {journal}
  {New Journal of Physics}\ }\textbf {\bibinfo {volume} {14}},\ \bibinfo
  {pages} {123022} (\bibinfo {year} {2012})}\BibitemShut {NoStop}%
\bibitem [{\citenamefont {Zhang}\ \emph {et~al.}(2013)\citenamefont {Zhang},
  \citenamefont {Chen},\ and\ \citenamefont {Zhang}}]{Zhang2011}%
  \BibitemOpen
  \bibfield  {author} {\bibinfo {author} {\bibfnamefont {Y.}~\bibnamefont
  {Zhang}}, \bibinfo {author} {\bibfnamefont {G.}~\bibnamefont {Chen}}, \ and\
  \bibinfo {author} {\bibfnamefont {C.}~\bibnamefont {Zhang}},\ }\href
  {\doibase 10.1038/srep01937} {\bibfield  {journal} {\bibinfo  {journal} {Sci.
  Rep.}\ }\textbf {\bibinfo {volume} {3}},\ \bibinfo {pages} {1937} (\bibinfo
  {year} {2013})}\BibitemShut {NoStop}%
\bibitem [{\citenamefont {D{\'o}ra}\ \emph {et~al.}(2012)\citenamefont
  {D{\'o}ra}, \citenamefont {Cayssol}, \citenamefont {Simon},\ and\
  \citenamefont {Moessner}}]{Balaz2012}%
  \BibitemOpen
  \bibfield  {author} {\bibinfo {author} {\bibfnamefont {B.}~\bibnamefont
  {D{\'o}ra}}, \bibinfo {author} {\bibfnamefont {J.}~\bibnamefont {Cayssol}},
  \bibinfo {author} {\bibfnamefont {F.}~\bibnamefont {Simon}}, \ and\ \bibinfo
  {author} {\bibfnamefont {R.}~\bibnamefont {Moessner}},\ }\href {\doibase
  10.1103/PhysRevLett.108.056602} {\bibfield  {journal} {\bibinfo  {journal}
  {Phys. Rev. Lett.}\ }\textbf {\bibinfo {volume} {108}},\ \bibinfo {pages}
  {056602} (\bibinfo {year} {2012})}\BibitemShut {NoStop}%
\bibitem [{\citenamefont {Struck}\ \emph {et~al.}(2012)\citenamefont {Struck},
  \citenamefont {{\"O}lschl{\"a}ger}, \citenamefont {Weinberg}, \citenamefont
  {Hauke}, \citenamefont {Simonet}, \citenamefont {Eckardt}, \citenamefont
  {Lewenstein}, \citenamefont {Sengstock},\ and\ \citenamefont
  {Windpassinger}}]{Struck2012}%
  \BibitemOpen
  \bibfield  {author} {\bibinfo {author} {\bibfnamefont {J.}~\bibnamefont
  {Struck}}, \bibinfo {author} {\bibfnamefont {C.}~\bibnamefont
  {{\"O}lschl{\"a}ger}}, \bibinfo {author} {\bibfnamefont {M.}~\bibnamefont
  {Weinberg}}, \bibinfo {author} {\bibfnamefont {P.}~\bibnamefont {Hauke}},
  \bibinfo {author} {\bibfnamefont {J.}~\bibnamefont {Simonet}}, \bibinfo
  {author} {\bibfnamefont {A.}~\bibnamefont {Eckardt}}, \bibinfo {author}
  {\bibfnamefont {M.}~\bibnamefont {Lewenstein}}, \bibinfo {author}
  {\bibfnamefont {K.}~\bibnamefont {Sengstock}}, \ and\ \bibinfo {author}
  {\bibfnamefont {P.}~\bibnamefont {Windpassinger}},\ }\href {\doibase
  10.1103/PhysRevLett.108.225304} {\bibfield  {journal} {\bibinfo  {journal}
  {Phys. Rev. Lett.}\ }\textbf {\bibinfo {volume} {108}},\ \bibinfo {pages}
  {225304} (\bibinfo {year} {2012})}\BibitemShut {NoStop}%
\bibitem [{\citenamefont {Bao}\ \emph {et~al.}(2003)\citenamefont {Bao},
  \citenamefont {Jaksch},\ and\ \citenamefont {Markowich}}]{Bao2003}%
  \BibitemOpen
  \bibfield  {author} {\bibinfo {author} {\bibfnamefont {W.}~\bibnamefont
  {Bao}}, \bibinfo {author} {\bibfnamefont {D.}~\bibnamefont {Jaksch}}, \ and\
  \bibinfo {author} {\bibfnamefont {P.~A.}\ \bibnamefont {Markowich}},\ }\href
  {\doibase 10.1016/S0021-9991(03)00102-5} {\bibfield  {journal} {\bibinfo
  {journal} {J. of Comp. Phys.}\ }\textbf {\bibinfo {volume} {187}},\ \bibinfo
  {pages} {318} (\bibinfo {year} {2003})}\BibitemShut {NoStop}%
\bibitem [{\citenamefont {Lignier}\ \emph {et~al.}(2007)\citenamefont
  {Lignier}, \citenamefont {Sias}, \citenamefont {Ciampini}, \citenamefont
  {Singh}, \citenamefont {Zenesini}, \citenamefont {Morsch},\ and\
  \citenamefont {Arimondo}}]{Lignier2007}%
  \BibitemOpen
  \bibfield  {author} {\bibinfo {author} {\bibfnamefont {H.}~\bibnamefont
  {Lignier}}, \bibinfo {author} {\bibfnamefont {C.}~\bibnamefont {Sias}},
  \bibinfo {author} {\bibfnamefont {D.}~\bibnamefont {Ciampini}}, \bibinfo
  {author} {\bibfnamefont {Y.}~\bibnamefont {Singh}}, \bibinfo {author}
  {\bibfnamefont {A.}~\bibnamefont {Zenesini}}, \bibinfo {author}
  {\bibfnamefont {O.}~\bibnamefont {Morsch}}, \ and\ \bibinfo {author}
  {\bibfnamefont {E.}~\bibnamefont {Arimondo}},\ }\href {\doibase
  10.1103/PhysRevLett.99.220403} {\bibfield  {journal} {\bibinfo  {journal}
  {Phys. Rev. Lett.}\ }\textbf {\bibinfo {volume} {99}},\ \bibinfo {pages}
  {220403} (\bibinfo {year} {2007})}\BibitemShut {NoStop}%
\end{thebibliography}%

\end{document}